\documentclass[prb,showpacs,preprintnumbers,amsmath,amssymb]{revtex4}
\usepackage{bm}
\usepackage{amsmath}
\usepackage[dvips]{graphicx}
\begin{document}
\bibliographystyle{apsrev}

\title{Superconducting Spiral Phase in the two-dimensional t-J model}

\author{Oleg P. Sushkov}
\email[E-mail:]{ sushkov@phys.unsw.edu.au} \affiliation{School of
Physics, University of New South Wales, Sydney 2052, Australia}
\author{Valeri N. Kotov }
\email[E-mail:]{ Valeri.Kotov@ipt.unil.ch}
\affiliation{Institute of Theoretical Physics, 
 University of Lausanne, 
CH-1015 Lausanne, Switzerland}

\date{\today}
\begin{abstract}
We analyse the $t-t'-t''-J$ model, relevant to the  superconducting cuprates.
By using  chiral perturbation theory we have determined  the ground state 
to be a spiral for small 
doping  $\delta \ll 1$ near  half filling.
 In this limit the solution does not contain any uncontrolled approximations.
We evaluate the spin-wave Green's functions and   address the issue of 
  stability of the  
spiral state, leading to   the  phase
diagram of the model. At $t'=t''=0$ the spiral state
is unstable  towards a local enhancement of the spiral pitch, and  the 
 nature of the true ground state remains unclear.
However, for values of $t'$ and $t''$ corresponding to real cuprates
the (1,0) spiral state is stabilized by  quantum fluctuations 
(``order from disorder'' effect).
We show that at   $\delta\approx 0.119$ the spiral is
commensurate with the lattice with a period  of 8 lattice spacings.
It is also demonstrated  that spin-wave mediated superconductivity
 develops in the spiral state and 
a lower limit for the superconducting gap is derived.
 Even though one cannot classify the gap symmetry according to the lattice
representations (s,p,d,...) since the symmetry of the lattice is spontaneously 
broken by the spiral, the gap always has  lines of
nodes along the $(1,\pm 1)$ directions.

\end{abstract}
  \maketitle

\section{Introduction}
The $t-J$ model has been suggested to describe the essential low-energy
physics of  the high-T$_c$ cuprates \cite{And0,Emery,ZR}.
Formation of spirals in the doped $t-J$ model was proposed by
Shraiman and Siggia \cite{ShSi}. They showed that  
the pitch of the spiral  is  proportional to the hole concentration
$\delta$, see also Ref. \cite{ShSi2}.
The idea that   the spiral state is the  ground state of
the  doped Heisenberg antiferromagnet attracted the  attention of theorists
\cite{Dom,Ng,Auer,IF,Mori,CM,Bruce,Manuel},  
however the question of stability of the state  remained controversial.
According to Refs. \cite{Dom,Auer,Mori} the spiral state is unstable toward
a local enhancement of the spiral pitch.
On the other hand the  analysis of Ref. \cite{IF} indicated  that 
the spiral state is stable.
A semiclassical analysis of stability of the (1,0) spiral state was performed
in Ref. \cite{ShSi2}. According to this analysis the state is marginal
(zero stiffness), which effectively  indicates that the state is unstable.
A complete  stability analysis of  spiral states in the Hubbard
model was performed in Ref. \cite{CM}  also using the  semiclassical
approximation. According to their analysis in the leading in powers of doping
approximation the (1,1) spiral is always unstable and the (1,0) spiral is 
always marginal in agreement with \cite{ShSi2}.
Recently the interest in the spiral state was renewed  because of
the strong experimental indications that at  small doping the cuprates
behave as spin glasses or even exhibit some kind of  magnetic ordering.
A summary of the available experimental data on one of
 the superconducting cuprates, La$_{2- \delta}$Sr$_{\delta}$CuO$_4$,  is given in  Ref. \cite{Jul}.
The data also  show that magnetic ordering and 
superconductivity coexist. 
The spin glass behavior is consistent with the spiral scenario:
since doping is not uniform the pitch of the spiral is varying
from point to point and hence on  large scales it
leads to  spin glass behavior \cite{HCS}.

In the present work we analyse the stability of  spiral states within the 
RPA approximation. The approximation is parametrically justified
for $\delta \ll 1$. In this part of the work on the technical side
we follow the approach developed by Igarashi and Fulde \cite{IF}.
However, contrary to them and in agreement with Refs. \cite{Dom,Auer}, 
we conclude that in  the ``pure'' $t-J$ model ($t'=t''=0$) the spiral state 
is unstable with respect to a local enhancement of the spiral pitch.
We find that relatively small values of $t'$ and $t''$ 
stabilize the spiral order, with a uniform hole distribution.
Our results are then consistent with numerical (DMRG)
results indicating that inhomogeneous (striped) phases disappear with the increase of
 further-neighbor hoppings 
\cite{White,Tohyama}.
In our approach the instability of the spiral
is closely related to the fact that the hole dispersion is almost
degenerate along the face of the magnetic Brillouin zone. As soon
as the degeneracy is sufficiently lifted by $t'$ and $t''$ the instability 
disappears, leading to a stable (1,0) spiral state. Within  the semiclasical approximation 
the effective stiffness of this state is zero in agreement with 
\cite{ShSi2,CM}.
However spin quantum fluctuations give rise to a nonzero positive
stiffness and hence favor stability via an ``order from disorder'' mechanism. 
For parameter values  corresponding to real cuprates, the pitch of the spiral
is proportional to doping, and at $\delta\approx 0.119$ the spiral becomes 
commensurate  with the lattice with period 8, in agreement with the 
experimental data of Tranquada {\it et al} \cite{Tranq,Tranq1}.

The possibility of s- and d-wave pairing between mobile holes due to
 a static 
distortion of the Neel background was pointed out in Ref. \cite{ShSi1}.
Such a distortion leads to an infinite set of very shallow two-hole 
 bound states \cite{KS}. In Ref. \cite{FKS} it was shown, 
under the  assumption 
that Neel order  is preserved under 
doping, that in the $t-J$ model there is  spin-wave mediated
superconducting pairing in all waves except  the  s-wave. The pairing
was found to be maximum in the d-wave channel. 
A numerical calculation  performed in Ref.\cite{BCDS}
under the same assumption showed that the pairing was  quite strong.
However, the picture of pairing \cite{FKS,BCDS} was not consistent because 
the staring point of the analysis was the  (unstable)  Neel background.
In the present paper, after proving the stability of the spiral state, 
we consider superconducting pairing on this state. 
Using an approach similar to that of Ref.\cite{FKS} we find 
a superconducting  pairing instability and
show that the gap always has lines of nodes along $(1,\pm 1)$ directions.
We estimate analytically the lower limit for the superconducting gap.
One cannot classify the gap according to the 
lattice symmetry representations  since the symmetry  is spontaneously 
broken by the spiral. 
 
In essence the method we use is  chiral perturbation theory 
which allows the treatment of strong
interactions and is exact in the long wave-length limit \cite{CP}.
The small parameter of the approach is doping near half filling, 
$\delta \ll 1$.
We cannot determine reliably  whether $\delta =0.15$, 
 $\delta =0.1$ or $\delta =0.05$ 
is sufficiently small to justify our calculations, but we claim
that at sufficiently small $\delta$ only the long range dynamics is  
important and the approach is parametrically justified.
The analysis of pairing within the chiral perturbation theory
by construction complies with the Adler relation \cite{Adler} and hence
with the argument presented by Schrieffer \cite{Sch}  concerning pairing via exchange of
Goldstone excitations.

The structure of the paper is as follows.
In Section \ref{Neel}  we calculate the single hole dispersion and the 
quasiparticle residue for different values of $t'$ and $t''$.
To do so we use the self consistent Born approximation. After that
we perform the RPA analysis and demonstrate  the instability of the Neel order, 
at arbitrary small hole concentrations, towards formation of spirals.
In Section III we consider the hole dispersion in different spiral states
and compare the total energies of the states.
The RPA analysis of stability of the (1,0) spiral state is performed in Section
IV, and in Section V we calculate the reduction of the 
spiral on-site magnetization due to doping.
Section VI is devoted to the superconducting pairing in spiral states. We 
present our conclusions in Section VII.

\section{Single hole properties and RPA proof of instability of the Neel order
upon doping}\label{Neel}

The Hamiltonian of the $t-t'-t''-J$ model is:
\begin{equation}
\label{H}
H=-t\sum_{\langle ij \rangle \sigma} c_{i\sigma}^{\dag}c_{j\sigma}
-t'\sum_{\langle ij_1 \rangle \sigma}
c_{i\sigma}^{\dag}c_{j_1\sigma}
-t''\sum_{\langle ij_2 \rangle \sigma} 
c_{i\sigma}^{\dag}c_{j_2\sigma}
+J \sum_{\langle ij \rangle \sigma} \left({\bf S}_i{\bf S}_j
-{1\over 4}n_in_j\right) \ .
\end{equation}
$c_{i \sigma}^{\dag}$ is the  creation operator of an electron with
spin $\sigma$ $(\sigma =\uparrow, \downarrow)$ at site $i$
of the two-dimensional square lattice,  $\langle ij \rangle$ represents
nearest neighbor sites, $\langle ij_1 \rangle$ - next nearest neighbors
(diagonal),
 and $\langle ij_2 \rangle$ represents next next
nearest sites.
 The spin operator is ${\bf S}_i={1\over 2}
c_{i \alpha}^{\dag} {\bf \sigma}_{\alpha \beta} c_{i \beta}$,
and the number density operator is 
$n_i=\sum_{\sigma}c_{i \sigma}^{\dag}c_{i \sigma}$.
The size of the exchange measured in two magnon Raman scattering
\cite{Tok,Grev} is $J=125meV$. 
Calculations of the
hopping matrix elements have been performed by Andersen {\it et al}
\cite{And}. They consider a  two-plane situation and the effective
matrix elements are slightly different for symmetric and
 antisymmetric combinations of orbitals between planes. After averaging
over these combinations we obtain: $t=386meV$, $t'=-105meV$,
$t''=86meV$. From now on we  set $J=1$. In  these units we have:
\begin{equation}
\label{ts}
t=3.1, \ \ \ t'=-0.8, \ \ \ t''=0.7
\end{equation}
An analysis of angle-resolved-photoemission spectra for  the insulating
copper oxide Sr$_2$CuO$_2$Cl$_2$
performed in Ref. \cite{SSEE} with the Hamiltonian (\ref{H}),(\ref{ts})
shows an excellent agreement with experiment for both the single-hole
dispersion and for the photoemission intensity. 
This analysis is based on the Self-consistent Born Approximation (SCBA)
\cite{Schmitt,Kane}. The approximation works very well due to the
absence of  single loop corrections to the hole-spin-wave vertex
\cite{Mart,Liu}. In the present calculation of
single-hole properties we follow the approach of Ref. \cite{SSEE}
and use  the SCBA.

It is well known that without doping (half filling) the Hamiltonian (\ref{H})
is equivalent to the 2D Heisenberg model which has 
antiferromagnetic (Neel) order.
There are two sublattices: sublattice $a$ with spin up
and sublattice $b$ with spin down.
Hence a hole created in the system has a pseudospin index $a$ or $b$.
The bare hole operator $d_i$ is defined so that $d_{i}^{\dag}
\propto c_{i \uparrow}$ on the $a$ sublattice and 
$\propto c_{i \downarrow}$ 
on the $b$ sublattice. In the momentum representation:
\begin{equation}
\label{d}
d_{{\bf k}a}^{\dag} =\sqrt{2\over {N (1/2+m)}}\sum_{i \in \uparrow}
c_{i \uparrow}e^{i{\bf k}{\bf r}_i}, \ \ \ \
d_{{\bf k}b}^{\dag} =\sqrt{2\over {N (1/2+m)}}\sum_{j \in \downarrow}
c_{j \downarrow}e^{i{\bf k}{\bf r}_j},
\end{equation}
where $N$ is the  number of sites and $m= |\langle 0|S_{iz}|0\rangle|\approx 0.3$ is
the average sublattice magnetization. The quasi-momentum
${\bf k}$ is limited to be  inside the magnetic
Brillouin zone: $\gamma_{\bf k}= {1\over 2}(\cos k_x + \cos k_y) \ge 0$.
The index $a$ ($b$) can be considered as pseudospin.
Rotational invariance is spontaneously broken, but nevertheless
the pseudospin gives the correct value of the spin $z$-projection:
$a$ corresponds to $S_z=-1/2$ and $b$ corresponds to $S_z=+1/2$.
The coefficients in (\ref{d}) provide the correct normalization:
\begin{equation}
\label{norm}
\langle 0|d_{{\bf k}\downarrow} d_{{\bf k}\downarrow}^{\dag}|0\rangle
={2\over {N (1/2+m)}}\sum_{i \in \uparrow}
\langle 0|c_{i \uparrow}^{\dag}c_{i \uparrow}|0\rangle
={1\over {1/2+m}}\langle 0|{1\over 2}+S_{iz}|0\rangle=1.
\end{equation}

For spin excitations the usual linear spin-wave theory is used
(see, e.g. the review paper\cite{Manousakis}).
Spin-wave excitations are described by operators
$\alpha_{\bf q}^{\dag}$ and $\beta_{\bf q}^{\dag}$ creating
spin waves with $S_z=-1$ and $S_z=+1$ respectively.
The momentum ${\bf q}$ is restricted inside the magnetic Brillouin zone. 
The operators are defined by the equations:
\begin{eqnarray}
\label{swt}
\sqrt{2\over N}\sum_{i \in \uparrow}S_i^+e^{-i{\bf q r_i}}&\approx&
u_{\bf q}\alpha_{\bf q}+v_{\bf q}\beta_{\bf -q}^{\dag},\\
\sqrt{2\over N}\sum_{j \in \downarrow}S_j^-e^{i{\bf q r_j}}&\approx&
v_{\bf q}\alpha_{\bf q}^{\dag}+u_{\bf q}\beta_{\bf -q}.\nonumber
\end{eqnarray}
The spin-wave dispersion
and the parameters of the Bogoliubov transformation diagonalizing the
spin-wave Hamiltonian are:
\begin{eqnarray}
\label{sw}
&&\omega_{\bf q}=2\sqrt{1-\gamma_{\bf q}^2},\nonumber \\
&&u_{\bf q}=\sqrt{{1\over{\omega_{\bf q}}}+{1\over 2}},\\
&&v_{\bf q}=-sign(\gamma_{\bf q})
\sqrt{{1\over{\omega_{\bf q}}}-{1\over 2}}.\nonumber
\end{eqnarray}

\begin{figure}[ht]
\centering
\includegraphics[height=160pt,keepaspectratio=true]{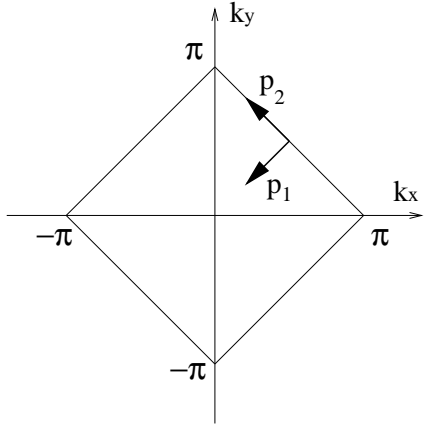}
\caption{\it Magnetic Brillouin zone.} 
\label{Fig1} 
\end{figure}

Hopping to a nearest neighbor site in the Hamiltonian (\ref{H}) leads to an
interaction of the hole with  the spin-waves:
\begin{equation}
\label{hsw}
H_{h,sw}=\sum_{\bf k,q}g_{\bf k,q} \left(
d_{{\bf k+q}a}^{\dag}d_{{\bf k}b}\alpha_{\bf q}+
d_{{\bf k+q}b}^{\dag}d_{{\bf k}a}\beta_{\bf q}+
H.c.\right),
\end{equation}
with the vertex $g_{\bf k,q}$ given by \cite{Mart,Liu}
\begin{eqnarray}
\label{g}
g_{\bf k,q} &\equiv& \langle 0|\alpha_{\bf q}d_{{\bf k}b}|H_t|
d_{{\bf k+q}a}^{\dag}|0\rangle
=4t\sqrt{2\over N}(\gamma_{\bf k}u_{\bf q}+
\gamma_{\bf k+q}v_{\bf q}).
\end{eqnarray}
The vertex $g_{\bf k,q}$ is independent of $t'$,
$t''$ because these parameters correspond to hopping inside
one sublattice. 
In agreement with Adler relation \cite{Adler} the vertex vanishes at
$q \to 0$.
We calculate the hole Green's function  using the SCBA
approximation \cite{Schmitt,Kane,Mart,Liu}.
This gives the quasiparticle dispersion $\epsilon_{\bf k}$ 
and the quasiparticle residue $Z_{\bf k}$. In the vicinity of 
the dispersion minima, ${\bf k}_0=(\pm\pi/2,\pm\pi/2)$, the quasiparticle 
residue is:
\begin{equation}
\label{Z}
Z_{\bf k} \approx  Z \equiv Z_{{\bf k}_0},
\end{equation}
and the dispersion can be approximated as
\begin{eqnarray}
\label{e}
\epsilon_{\bf k} \approx  const+
\beta_1\gamma^2_{\bf k}+\beta_2(\gamma^-_{\bf k})^2\approx
const + \beta_1\frac{p_1^2}{2}+\beta_2\frac{p_2^2}{2} \to
\beta_1\frac{p_1^2}{2}+\beta_2\frac{p_2^2}{2}\ .
\end{eqnarray}
Here $\gamma^-_{\bf k}={1\over 2}(\cos k_x - \cos k_y)$,  
 ${\bf p}={\bf k}-{\bf k}_0$, the component $p_1$ is orthogonal to the
face of the Brillouin zone and the component $p_2$ is parallel to the face,
see Fig.\ref{Fig1}.
The results of our calculations at $t=3.1$ and for the second neighbor 
 hoppings within the intervals $-1<t'<0$ and 
$0 < t''<1$ can be fitted as:
\begin{eqnarray}
\label{ttt}
t&=&3.1 \ ,\nonumber\\
\beta_1&=&1.96 + 1.15t'+0.06t'^2 
+2.70t''+0.53t''^2+0.50t't''\ ,\nonumber\\
\beta_2&=&0.30 - 1.33t'-0.19t'^2+2.80t''+1.06t''^2-0.14t't''\ ,\nonumber\\
Z&=&0.29 + 0.055t'+0.195t''\ .
\end{eqnarray}
\noindent
These formulae agree with the results of Refs. \cite{Mart,Liu} at $t'=t''=0$,
as well as  with Ref. \cite{SSEE} for $t'$, $t''$ given by (\ref{ts}).
The quasiparticle-spin-wave interaction differs from the bare
interaction (\ref{hsw}) only
by the presence of the quasiparticle residues and reads:
\begin{equation}
\label{qpsw}
H_{qp,sw}=\sum_{\bf k,q}M_{\bf q} \left(
h_{{\bf k+q}a}^{\dag}h_{{\bf k}b}\alpha_{\bf q}+
h_{{\bf k+q}b}^{\dag}h_{{\bf k}a}\beta_{\bf q}+
H.c.\right),
\end{equation}
where $h^{\dag}_{{\bf k}a}$ and $h^{\dag}_{{\bf k}b}$
are quasihole creation operators, and the  vertex $M_{\bf q}$ has the form
(we set $N=1$ for convenience):
\begin{eqnarray}
\label{g1}
M_{\bf q} =\sqrt{Z_{\bf k}Z_{{\bf k}+{\bf q}}}g_{\bf k,q}
\approx -2^{7/4}Zt\frac{q_1}{\sqrt{q}}\ .
\end{eqnarray}
Here $q_1$ is  the component of ${\bf q}$ orthogonal to the face of the
Brillouin zone, and we assume that ${\bf k}\approx {\bf k}_0$ and $q \ll 1$.

\begin{figure}[ht]
\centering
\includegraphics[height=70pt,keepaspectratio=true]{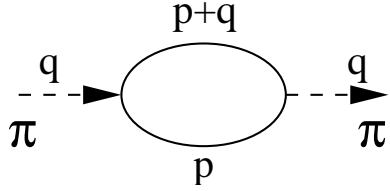}
\caption{\it $\pi_{\bf q}$-spin-wave polarization operator} 
\label{Fig2} 
\end{figure}

\begin{figure}[ht]
\centering
\includegraphics[height=170pt,keepaspectratio=true]{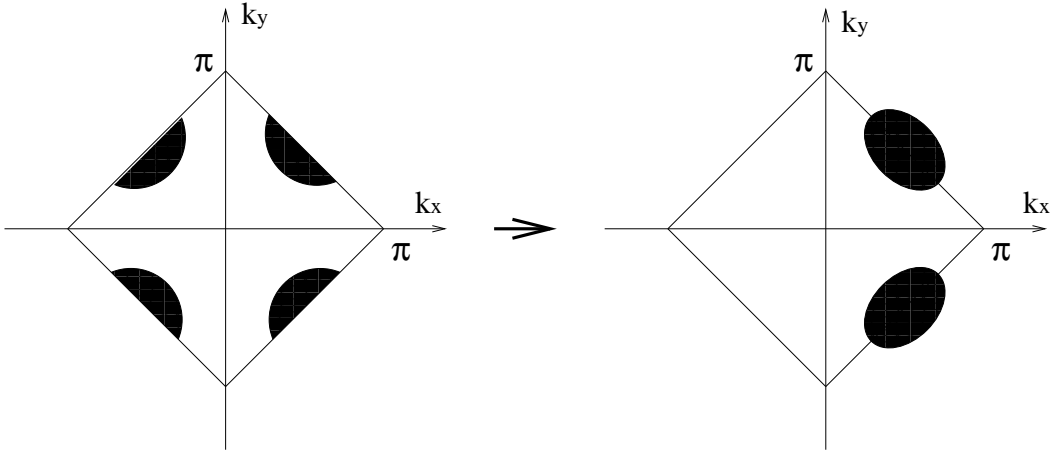}
\caption{\it Hole pockets and Fermi surface for the Neel state
and for the spiral state with ${\bf Q}\propto (1,0)$.} 
\label{Fig3} 
\end{figure}

It is convenient to introduce the fields $\pi_{\bf q}$ and $\lambda_{\bf q}$
instead of $\alpha_{\bf q}$ and $\beta_{\bf q}$:  
\begin{eqnarray}
\label{pi}
\pi_{\bf q}&=&\alpha_{\bf q}-\beta^{\dag}_{\bf -q} \ ,\nonumber\\
\lambda_{\bf q}&=&\alpha_{\bf q}+\beta^{\dag}_{\bf -q} \ .
\end{eqnarray}
The Green's functions of these fields are defined as:
\begin{eqnarray}
\label{D}
D_{\pi n}(\omega,{\bf q})&=&-i\int_{-\infty}^{+\infty}dte^{i\omega t}
\langle T[\pi_{\bf q}(t)\pi^{\dag}_{\bf q}(0)]\rangle \ ,\nonumber\\
D_{\lambda n}(\omega,{\bf q})&=&-i\int_{-\infty}^{+\infty}dte^{i\omega t}
\langle T[\lambda_{\bf q}(t)\lambda^{\dag}_{\bf q}(0)]\rangle \ .
\end{eqnarray}
In the absence of interactions they are: 
\begin{equation}
\label{D0}
D^{(0)}_{\pi n}(\omega,{\bf q})=D^{(0)}_{\lambda n}(\omega,{\bf q})
=\frac{2\omega_{\bf q}}{\omega^2-\omega_{\bf q}^2+i0}\ .
\end{equation}
The index $n$ in (\ref{D}), (\ref{D0}) labels them as 
normal Green's functions.
According to Eq.(\ref{qpsw}) the field $\pi_{\bf q}$
interacts with  the quasiholes:
\begin{equation}
\label{qppi}
H_{qp,\pi}=\sum_{\bf k,q}M_{\bf q} \left(
h_{{\bf k+q}a}^{\dag}h_{{\bf k}b}\pi_{\bf q}
+H.c.\right) \ ,
\end{equation}
while the field $\lambda_{\bf q}$ remains idle with respect to this
interaction.
The interaction (\ref{qppi}) generates a loop correction to the spin-wave 
Green's function shown in Fig. \ref{Fig2}.
\noindent
The holes are fermions and  occupy four half pockets or two full pockets 
as  shown in Fig.\ref{Fig3}.
\noindent
At a given momentum there are two states with different pseudospin,
so the Fermi energy $\epsilon_F$ and the total  Fermi motion
energy $E_F$ (per lattice site)  read:
\begin{eqnarray}
\label{ef1}
\epsilon_F&=&\frac{\pi}{2}\sqrt{\beta_1\beta_2}\delta\ ,\nonumber\\
E_F&=&\frac{\pi}{4}\sqrt{\beta_1\beta_2}\delta^2\ , 
\end{eqnarray}
where $\delta \ll 1$ is the hole concentration.
The calculation of the polarization operator Fig. \ref{Fig2} is straightforward
and gives:
\begin{eqnarray}
\label{pol1}
P(\omega=i\xi,{\bf q})&=&\sum_{pockets}M_{\bf q}^2\sum_{\bf p}
 \frac{2(\epsilon_{\bf p}-\epsilon_{\bf p+q})n_{\bf p}(1-n_{\bf p+q})}
{\xi^2+(\epsilon_{\bf p+q}-\epsilon_{\bf p})^2}
=-\frac{2^{5/2}Z^2t^2}{\pi\sqrt{\beta_1\beta_2}}\ \frac{1}{q}
\sum_{pockets}q_1^2F(\xi,{\bf q}) \ ,\\
F(\xi,{\bf q})&=&\frac{1}{t^2}
\left[t^2-2Re\sqrt{(t^2/2+i\xi)^2-2\epsilon_Ft^2}\right] \ ,
\ \ F(0,0)=1  \ . \nonumber
\end{eqnarray}
Here $t^2=\beta_1q_1^2+\beta_2q_2^2$ and
$n_{\bf p}$ is the Fermi distribution function. 
Re stays for real part. Summation over two full
pockets shown on the right side of Fig. \ref{Fig3} has to be performed.
Note that what is $q_1$ for one pocket is $q_2$ for the other, and vice versa.
At $\omega=0$ and $q \ll p_F\sim \sqrt{\delta}$ the polarization operator
(\ref{pol1}) is particularly simple:
\begin{equation}
\label{pol2}
P(0,{\bf q})= -\frac{2^{5/2}Z^2t^2}{\pi\sqrt{\beta_1\beta_2}}\ q
\end{equation}
The polarization operator is always proportional to $q$, as a  
direct consequence of  Goldstone's theorem.
A crucial point is that the polarization operator (\ref{pol2})
is independent of $\delta$.
Taking into account the one-loop diagram Fig. \ref{Fig2} the $\pi$-magnon Green's function
(\ref{D}) becomes:
\begin{equation}
\label{D1}
D_{\pi n}(\omega,{\bf q})=\frac{2\omega_{\bf q}}{\omega^2-\omega_{\bf q}^2
-2\omega_{\bf q}P(\omega,{\bf q})+i0}\ .
\end{equation}
The stability of the system requires that all poles of the Green's function
(\ref{D1}) are at positive $\omega^2$. Using Eqs. (\ref{pol2}) and
(\ref{ttt}) one can see that this criterion is  violated  for  
$q \ll p_F\sim \sqrt{\delta}$ at arbitrary small doping, because (cf. with
Refs. \cite{SF,CM})
\begin{equation}
\label{st1}
\omega_{\bf q}^2+2\omega_{\bf q}P(\omega=0,{\bf q})
=2q^2\left(1-\frac{8Z^2t^2}{\pi\sqrt{\beta_1\beta_2}}\right)<0 \ .
\end{equation}
This result signals an instability towards formation of spirals.

In the presented RPA proof of the Neel state instability we have used
the following approximations:
(1.) The single hole properties (dispersion and quasiparticle residue) have been 
calculated without account of other holes.
(2.) The incoherent part of the hole Green's function has been neglected.
(3.) Interactions between quasiholes have been neglected.
These approximations are {\it parametrically} justified at sufficiently
small $\delta$ since (\ref{pol2}) is independent of $\delta$.
We observe that  doping influences the behavior of the 
system only at momenta $q\sim \sqrt{\delta}\ll 1$. At the same time all 
integrals in the SCBA are convergent at large momenta, $q\sim 1$.
Consequently doping gives a negligible correction ($\propto \delta$) to the
dispersion. The same is true for the incoherent part of the hole Green's 
function and for the hole-hole interaction. The leading correction from
these effects is $\propto \delta\ln\delta$, see Ref. \cite{MS}, and hence
it is also negligible for sufficiently small $\delta$.

\section{Spiral states, hole dispersion and total energy}
\begin{figure}[ht]
\centering
\includegraphics[height=170pt,keepaspectratio=true]{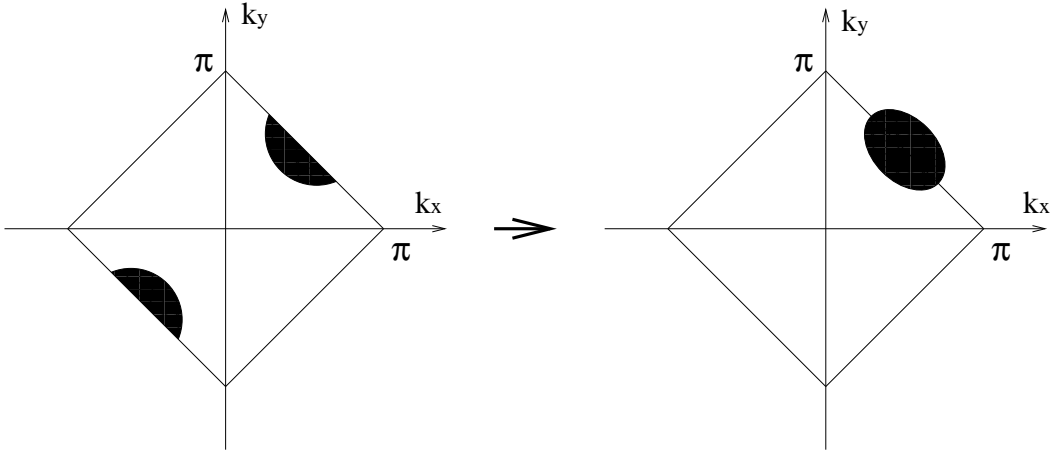}
\caption{\it Hole pockets and Fermi surface for the spiral state with
${\bf Q}\propto (1,1)$.} 
\label{Fig4} 
\end{figure}

\begin{figure}[ht]
\centering
\vspace{25pt}
\includegraphics[height=200pt,keepaspectratio=true]{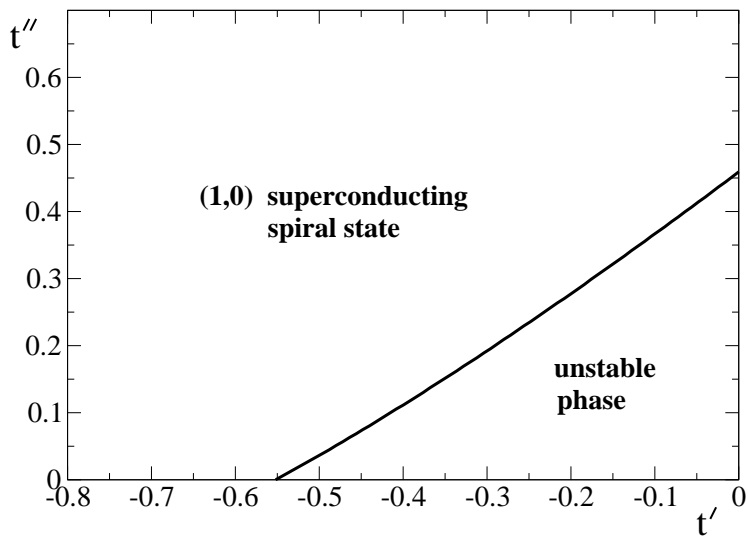}
\vspace{10pt}
\caption{\it Phase diagram of the $t-t'-t''-J$ model at
$t=3.1$ and small doping,  $\delta \ll 1$. 
The bottom right corner corresponds to the 
pure $t-J$ model, $t'=t''=0$. The top left corner corresponds to parameters
from Ref.\cite{And},  Eq. (\ref{ts}).
 The  region of stability of the superconducting (1,0) spiral phase is shown.} 
\label{Fig5} 
\end{figure}

In a spiral state there are still two sublattices, 
sublattice $a$ and sublattice $b$,
but the spin at every site $j$ of each sublattice is rotated by an angle
$\theta_j$:
\begin{eqnarray}
\label{spir}
|\psi_j\rangle&=&e^{i\theta_j{\bf n}\cdot{\bf \sigma}/2}|\uparrow\rangle \ \ \
if \ \  j \in \ a \ sublattice \ ,\nonumber\\
|\psi_j\rangle&=&e^{i\theta_j{\bf n}\cdot{\bf \sigma}/2}
|\downarrow\rangle \ \ \
if \ \  j \in \ b \ sublattice\ ,\nonumber\\
\theta_j&=&{\bf Q}\cdot {\bf r}_j \ .
\end{eqnarray} 
Here ${\bf Q}\ll 1$ is the  vector of the spiral, ${\bf \sigma}$ is the vector of Pauli matrices,
and ${\bf n}$ is an axis of rotation. The direction of ${\bf n}$ can be arbitrary
in the plane orthogonal to the spin $|\uparrow\rangle$ in Eq. (\ref{spir}),
${\bf n}=(n_1,n_2,0)=(\cos\alpha,\sin\alpha,0)$.
To first order in $Q$ the small rotation (\ref{spir}) does not influence
the spin-wave dispersion and the hole-spin-wave interaction considered
in the previous section. The only effect which appears at first order in  $Q$
is the possibility for a hole to hop between nearest neighbor sites
of the lattice. Using (\ref{H}) and (\ref{spir}) one can easily find the
Hamiltonian describing this hopping:
\begin{equation}
\label{Hh}
H_Q=-t\sum_{\xi,i\in a}\sin\left(\frac{\bf Q\cdot \xi}{2}\right)
\left[ie^{-i\alpha}d^{\dag}_{i+\xi,b}d_{ia}-
ie^{i\alpha}d^{\dag}_{ia}d_{i+\xi,b}
\right] \ .
\end{equation}
Here $\xi$ is a lattice vector directed from a given site $i\in a$ to the 
nearest neighbor $i+\xi \in b$. One can rewrite (\ref{Hh}) in momentum
representation using the quasihole operators $h_{\bf k}$:
\begin{eqnarray}
\label{Hhh}
H_Q&=&-t\sum_{\bf k}(Q_x\sin k_x + Q_y\sin k_y)
\left(e^{-i\alpha}d^{\dag}_{{\bf k}b}d_{{\bf k}a}+
e^{i\alpha}d^{\dag}_{{\bf k}a}d_{{\bf k}b}
\right) \nonumber \\
&\to&
-Z_{\bf k}t\sum_{\bf k}(Q_x\sin k_x + Q_y\sin k_y)
\left(e^{-i\alpha}h^{\dag}_{{\bf k}b}h_{{\bf k}a}+
e^{i\alpha}h^{\dag}_{{\bf k}a}h_{{\bf k}b}
\right) \ .
\end{eqnarray}
The quasiparticle residue $Z_{\bf k}$ is exactly the
same as the one found in the previous section by the SCBA.
Indeed,  the effects considered in the present section
are relevant to the long range dynamics, so the corresponding
momenta are very small, $Q \propto \delta \ll 1$. On the other hand
all the integrals in the SCBA are convergent at large momenta and therefore 
 not sensitive to the long range dynamics.
The axes x and y coincide with the crystal axes. Note that the axes 1 and 2
used to define the phase $\alpha$ are not related to the crystal axes.
The Hamiltonian (\ref{Hhh}) mixes the states $h^{\dag}_{{\bf k}a}$ and
$h^{\dag}_{{\bf k}b}$. Therefore the effective Hamiltonian matrix for a given 
${\bf k}$ takes the form:
\begin{equation}
\label{Hab} 
H_{eff}= \left(\begin{matrix}
\epsilon_{\bf k} \\
-Z_{\bf k} t e^{i\alpha} [Q_x\sin k_x + Q_y\sin k_y]
\end{matrix}
\begin{matrix}
-Z_{\bf k} t e^{-i\alpha} [Q_x\sin k_x + Q_y\sin k_y]\\
 \epsilon_{\bf k}
\end{matrix}\right)
\quad ,
\end{equation}
where $\epsilon_{\bf k}$ is found in the previous section.
In the vicinity of the points
${\bf k}_0=(\pm\pi/2,\pm\pi/2)$ one can use Eqs. (\ref{Z}) and (\ref{e})
and hence $H_{eff}$ takes the following form:
\begin{equation}
\label{Hab1} 
H_{eff}= \left(\begin{matrix}
\epsilon_{\bf k} \\
\sqrt{2}Z t e^{i\alpha} Q_1
\end{matrix}
\begin{matrix}
\sqrt{2}Z t e^{-i\alpha} Q_1\\
 \epsilon_{\bf k}
\end{matrix}\right)
\quad ,
\end{equation}
where $Q_1$ is the component of ${\bf Q}$ orthogonal to the corresponding
face of the magnetic Brillouin zone.
We denote by $\psi_{{\bf k}}^{\dag}$ and $\varphi_{{\bf k}}^{\dag}$
the creation operators that  diagonalize this
Hamiltonian
\begin{eqnarray}
\label{psis}
\psi_{{\bf k}}^{\dag}&=&\frac{1}{\sqrt{2}}\left(
h_{{\bf k}a}^{\dag}-e^{-i\mu}h_{{\bf k}b}^{\dag}
\right) \ ,\nonumber\\
\varphi_{{\bf k}}^{\dag}&=&\frac{1}{\sqrt{2}}\left(
h_{{\bf k}a}^{\dag}+e^{-i\mu}h_{{\bf k}b}^{\dag}
\right) \ ,
\end{eqnarray}
where $e^{i\mu}=\frac{Q_1}{|Q_1|}e^{i\alpha}$.
The corresponding dispersions are:
\begin{eqnarray}
\label{es}
\epsilon_{{\bf k}-}&=\epsilon_{\bf k}-\sqrt{2}Z t |Q_1| \ ,\nonumber\\
\epsilon_{{\bf k}+}&=\epsilon_{\bf k}+\sqrt{2}Z t |Q_1| \ .
\end{eqnarray}
We will see that $|Q_1|$ is always large enough, so  only the states
corresponding to the $\epsilon_{{\bf k}-}$ branch ($\psi$-band) 
are filled by holes. 
According to Eq. (\ref{es}) the total kinetic energy gain due to the spiral 
is $\Delta E_{kin}= -\sqrt{2}Z t |Q_1|\delta$ per lattice site. 
On the other hand the spiral increases the magnetic energy.
The variation of the magnetic energy per lattice site is
$\Delta E_{magn}=\frac{1}{2}\rho_s Q^2$, where 
$\rho_s =Z_{\rho}/4\approx 0.18$ is the
spin stiffness,   and $Z_{\rho}\approx 0.72$ is the
renormalization factor due to higher $1/S$-corrections \cite{SZ}.
Finally, the Fermi motion of holes $E_F\propto \delta^2$, 
also contribute to the total energy. Altogether the total energy of the 
spiral state  (per site) with respect to the undoped antiferromagnet is:
\begin{equation}
\label{dE}
E = \frac{1}{2}\rho_s Q^2-\sqrt{2}Z t |Q_1|\delta +E_F  \ .
\end{equation}
There are two possibilities to minimize the total energy: for a spiral directed 
along a diagonal of the lattice ${\bf Q}\propto (1,1)$, and for a spiral 
directed along a crystal axis of the lattice ${\bf Q}\propto(1,0)$.
For the diagonal spiral the effective Fermi surface consists of two
half pockets or  one whole pocket, see Fig. \ref{Fig4}.
\noindent
The hole states do not have a   pseudospin index and hence
the Fermi energy $\epsilon_F$ and the total Fermi motion energy $E_F$ 
(per site) in this case (${\bf Q}\propto (1,1)$) read:
\begin{eqnarray}
\label{ef11}
\epsilon_F&=&2\pi\sqrt{\beta_1\beta_2}\delta\ ,\nonumber\\
E_F&=&\pi\sqrt{\beta_1\beta_2}\delta^2 \ .
\end{eqnarray}
For the spiral along a crystal axis the effective Fermi surface consists of 
four half pockets or  two whole pockets, similarly to the 
Neel case, see Fig. \ref{Fig3}. However  the hole  states in a spiral
do not carry a pseudospin index and hence the Fermi energy and the total Fermi 
motion energy $E_F$ in this case (${\bf Q}\propto (1,0)$) are:
\begin{eqnarray}
\label{ef10}
\epsilon_F&=&\pi\sqrt{\beta_1\beta_2}\delta\ ,\nonumber\\
E_F&=&\frac{\pi}{2}\sqrt{\beta_1\beta_2}\delta^2 \ .
\end{eqnarray}
Using Eqs. (\ref{dE}), (\ref{ef11}), and (\ref{ef10})
we find the following expressions for the energy of the spiral state 
relative to the energy of the undoped antiferromagnet: 
\begin{eqnarray}
\label{dE1}
{\bf Q}\propto (1,1): & \ \ \ & 
E_{(1,1)} = \frac{1}{2}\rho_s Q^2-\sqrt{2}Z t Q \delta +
\pi\sqrt{\beta_1\beta_2}\delta^2 \ ,\nonumber\\
{\bf Q}\propto (1,0): & \ \ \ & 
E_{(1,0)} = \frac{1}{2}\rho_s Q^2-Z t Q \delta +
\frac{\pi}{2}\sqrt{\beta_1\beta_2}\delta^2 \ .
\end{eqnarray}
Minimizing $ E$ with respect to $Q$ one finds:
\begin{eqnarray}
\label{dE2}
{\bf Q}\propto (1,1): & \ \ \ & 
Q=\frac{\sqrt{2}Zt}{\rho_s}\delta \ , \ \ \ 
E_{(1,1)} = \left\{-\frac{Z^2 t^2}{\rho_s} +
\pi\sqrt{\beta_1\beta_2}\right\}\delta^2 \ ,\nonumber\\
{\bf Q}\propto (1,0): & \ \ \ & 
Q=\frac{Zt}{\rho_s}\delta \ , \ \ \ 
E_{(1,0)} = \left\{-\frac{Z^2 t^2}{2\rho_s} +
\frac{\pi}{2}\sqrt{\beta_1\beta_2}\right\}\delta^2 \ .
\end{eqnarray}
Eqs. (\ref{dE1}) and (\ref{dE2}) agree with Ref.\cite{IF}, see also Ref. 
\cite{Eder}.
We see that $E_{(1,1)}=2E_{(1,0)}$, so if these energies are positive then
the state (1,0) has lower energy than the (1,1) state. 
It should be noted that the  energy $E_{(1,0)}$ is always lower than the energy of the
doped Neel state (i.e. the difference between $E_{(1,0)}$ and Eq.(\ref{ef1}) is always
 negative). 
When $E_{(1,1)}$ is negative then this state has lower energy. However, in
this case the corresponding compressibility 
$\frac{\partial^2E_{(1,1)}}{\partial\delta^2}$ is also negative
which implies an instability towards a state with an inhomogeneous
charge distribution.
 Consequently  the (1,1) spiral state is always unstable in agreement with 
the conclusions of Refs. \cite{Dom,Auer,CM}. On the other hand the comressibility
 of the (1,0) state is always positive (in 
agreement with Ref. \cite{CM}).
Using the results of our  SCBA calculations, summarized in Eqs. (\ref{ttt}),  
as well as Eqs. (\ref{dE2}), one can easily find the region where the state
(1,0) is realised.  Our results are summarized in  Fig. \ref{Fig5}. 
The energy $E_{(1,0)}$ vanishes along the solid line, and  the spiral state
is unstable in the right bottom corner of the phase diagram.
To complete the analysis of stability of the (1,0) spiral state we must
calculate the spin-wave polarization operator (``transverse stiffness'').
We leave this analysis to the next section.
\noindent

For values of $t$, $t'$ and $t''$ from Ref.\cite{And}, Eq. (\ref{ts}), the 
$(1,0)$ spiral is stable.
At this point  $Z=0.38$,  Eq.(\ref{ttt}),  and hence according to 
Eq. (\ref{dE2}) the magnitude of the spiral vector is:
\begin{equation}
\label{qd}
Q=6.58\delta \ , \ \ \ \ at \ t=3.1,  \ t'=-0.8,  \ t''=0.7 
\end{equation}
The spiral is commensurate with the lattice if $Qn=\pi$, where $n$
is an integer number. The corresponding hole concentration
$\delta_n= \pi/(6.58 n)$. This results in an effective antiferromagnetic
structure with period $\Delta r$. The period is $\Delta r =n$ for odd values of 
$n$, and $\Delta r =2n$ for even values of $n$.
 In  Table 1 we present the values of $\delta_n$  
and the corresponding periods of the commensurate antiferromagnetic structure for 
several values of $n$.

\begin{figure}[ht]
\centering
\includegraphics[height=70pt,keepaspectratio=true]{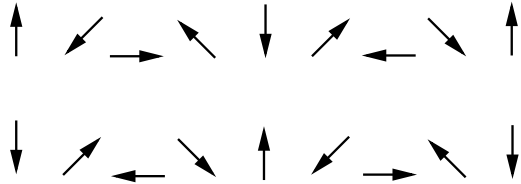}
\caption{\it (1,0) Spiral state  for $\delta = 0.119$, corresponding to
 $Q=\pi/4$.}
\label{45spiral}
\end{figure}

\vspace{0.8cm}
\begin{center}
\begin{tabular}{cccccc}\hline
$n$ \ \ \ \ & \ \ \ \ 3 \ \ \ \  & \ \ \ \ 4 \ \ \ \ & \ \ \ \ 5 \ \ \ \ & \ \ \ \ 6 \ \ \ \ \\
\hline
$\delta_n$ & 0.159 & 0.119 & 0.095 & 0.079\\ \hline
$\Delta r$ &  3 & 8 & 5 & 12 \\ \hline
\end{tabular}\\
\vspace{20pt}
Table I. {\it Values of the hole concentration $\delta_n$ at which a  
commensurate antiferromagnetic structure with period $\Delta r$  
is established.}
\vspace{10pt}
\end{center}
Notice that for $\delta_n=0.119$ the period is 8, in agreement with experimental 
data \cite{Tranq}. The spin structure at this  doping value is shown in Fig.
\ref{45spiral}. In section VII we discuss further the possible connections 
of our results to the physics of the superconducting cuprates.

\section{Spin-wave Green's function in the spiral state and RPA stability analysis 
of the state.}
\begin{figure}[ht]
\centering
\includegraphics[height=150pt,keepaspectratio=true]{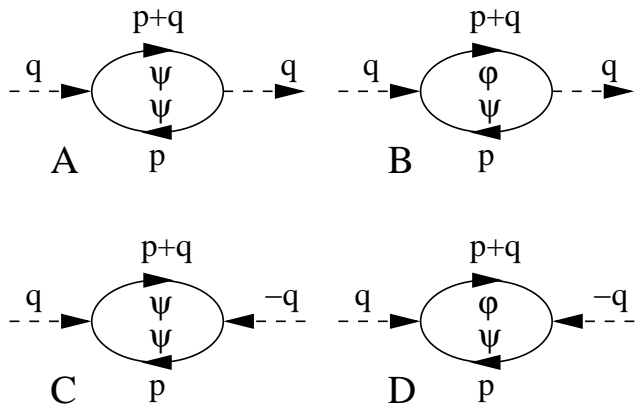}
\caption{\it Contributions to $\pi$-spin-wave polarization operators.
A and C describe transitions within the  $\psi$-band, and B and D
describe transitions between  the $\psi$- and $\varphi$-bands.} 
\label{Fig6} 
\end{figure}

The stability analysis of the (1,0) spiral state requires a calculation
of the spin-wave Green's functions (\ref{D}), similar to the calculation 
performed in Section \ref{Neel} for the Neel state. 
The spin wave $\lambda_{\bf q}$,  Eq. (\ref{pi}), remains unchanged since it 
does not interact with the holes.
The interaction of the spin wave $\pi_{\bf q}$ with the holes is given by 
Eq. (\ref{qppi}), but one has to take into account the fact that the
 correct quasiparticle 
states are represented by the operators $\psi_{{\bf k}}$ and $\varphi_{{\bf k}}$
instead of $h_{{\bf k}a}$ and $h_{{\bf k}b}$,  Eq. (\ref{psis}).
The diagrams shown in Fig. \ref{Fig6} contribute to the $\pi$-spin-wave 
polarization operators.
\noindent
The diagrams Fig. \ref{Fig6} A and B contribute to the normal polarization 
operator.
The filled state is always $\psi_{{\bf k}}$, however the excited state in the
loop can be either $\psi_{{\bf k+q}}$ or $\varphi_{{\bf k+q}}$, see Eq. (\ref{psis}), 
 providing the difference between the diagrams Fig. \ref{Fig6} A and B.
Since pseudospin is not conserved there is also an anomalous
polarization operator. The contributions to the anomalous operator are
given by the  diagrams Fig. \ref{Fig6} C and D.
The anomalous polarization operator gives rise to the anomalous
$\pi$-spin-wave Green's function defined as:
\begin{equation}
\label{Da}
D_{\pi a}(\omega,{\bf q})=-i\int_{-\infty}^{+\infty}dte^{i\omega t}
\langle T[\pi_{\bf -q}^{\dag}(t)\pi^{\dag}_{\bf q}(0)]\rangle \ .
\end{equation}
A straightforward calculation
 of the polarization operators $P_A$ and $P_C$, taking into account
 transitions within the
same subband, gives (we set $\omega =i\xi$ from now on): 
\begin{eqnarray}
\label{polAC}
P_A(i\xi,{\bf q})=-e^{2i\mu}P_C(i\xi,{\bf q})
&=&\sum_{pockets}\frac{M_{\bf q}^2}{4}\sum_{\bf p}
 \frac{2(\epsilon_{\bf p}-\epsilon_{\bf p+q})n_{\bf p}(1-n_{\bf p+q})}
{\xi^2+(\epsilon_{\bf p+q}-\epsilon_{\bf p})^2}
=-\frac{\sqrt{2}Z^2t^2}{\pi\sqrt{\beta_1\beta_2}}\ \frac{1}{q}
\sum_{pockets}q_1^2F(\xi,{\bf q}) \ .
\end{eqnarray}
The function $F(\xi,{\bf q})$ is defined in Eq. (\ref{pol1}).
Eq. (\ref{polAC}) differs from (\ref{pol1}) by the coefficient
1/4 only. 
The summation has to be performed over two full pockets of the
$(1,0)$ spiral state,  Fig. \ref{Fig3}.
We remind the reader that $q_1$ is the component orthogonal to the face of
the Brillouin zone.
Next we evaluate the  polarization operators $P_B$ and $P_D$ 
describing transitions between
subbands: 
\begin{eqnarray}
\label{polBD}
P_B(i\xi,{\bf q})=e^{2i\mu}P_D(i\xi,{\bf q})
&=&\sum_{pockets}\frac{M_{\bf q}^2}{4}\sum_{\bf p}
 \frac{2(\epsilon_{\bf p}-\epsilon_{\bf p+q}-\Delta)
n_{\bf p}}{\xi^2+(\epsilon_{\bf p+q}-\epsilon_{\bf p}+\Delta)^2}
=-\frac{\sqrt{2}Z^2t^2}{\pi\sqrt{\beta_1\beta_2}}\ \frac{1}{q}
\sum_{pockets}q_1^2\Phi(\xi,{\bf q}) \ ,\\
\Phi(\xi,{\bf q})&=&\frac{1}{t^2}
\left[t^2+2\Delta-2Re\sqrt{(t^2/2+\Delta+i\xi)^2-2\epsilon_Ft^2}\right] \ ,
\ \ \Phi(0,0)=\frac{2\epsilon_F}{\Delta} \ ,
\nonumber
\end{eqnarray}
where $\Delta=2\sqrt{2}Z t |Q_1|$ is  the splitting between the subbands,
see Eq. (\ref{es}). Taking  into account   Eq. (\ref{dE2}) 
for the (1,0) state we obtain:
\begin{eqnarray}
\label{dd}
\Delta&=& \frac{2Z^2 t^2}{\rho_s}\delta \approx 15\delta  \ .
\end{eqnarray}
According to Fig. \ref{Fig6} and Eqs. (\ref{polAC}), (\ref{polBD})
the normal  and the  anomalous polarization operators are:
\begin{eqnarray}
\label{pn}
P_n(i\xi,{\bf q})&=&
P_A(i\xi,{\bf q})+P_B(i\xi,{\bf q})
=-\frac{\sqrt{2}Z^2t^2}{\pi\sqrt{\beta_1\beta_2}}\ \frac{1}{q}
\sum_{pockets}q_1^2\left[F(\xi,{\bf q})+\Phi(\xi,{\bf q})\right] \ ,\nonumber\\
P_a(i\xi,{\bf q})&=&
P_C(i\xi,{\bf q})+P_D(i\xi,{\bf q})
=-\frac{\sqrt{2}Z^2t^2}{\pi\sqrt{\beta_1\beta_2}}\ \frac{e^{-2i\mu}}{q}
\sum_{pockets}q_1^2\left[-F(\xi,{\bf q})+\Phi(\xi,{\bf q})\right] \ .
\end{eqnarray}
The corresponding  Dyson's equations for  the Green's functions read:
\begin{eqnarray}
\label{Dy}
D_{\pi n }(q)&=&D_{\pi n }^{(0)}(q)+D_{\pi n}^{(0)}(q)P_n(q)D_{\pi n}(q)
+D_{\pi n}^{(0)}(q)P_a(q)D_{\pi a}(q) \ ,
\nonumber\\
D_{\pi a}(q)&=&D_{\pi n}^{(0)}(-q)P_n(-q)D_{\pi a}(q)+D_{\pi n}^{(0)}(-q)P^{*}_a(q)D_{\pi n}(q) 
\ ,
\end{eqnarray}
with $D_{\pi n}^{(0)}$ given by Eq. (\ref{D0}). By solving these equations
we obtain  
 the normal and the
anomalous Green's functions of the $\pi$-field:
\begin{eqnarray}
\label{Dna}
D_{\pi n}(\omega=i\xi,{\bf q})&=&
\frac{-2\omega_{\bf q}\left[\xi^2+\omega_{\bf q}^2+2\omega_{\bf q}P_n(i\xi,{\bf q})
\right]}
{\left[\xi^2+\omega_{\bf q}^2+2\omega_{\bf q}P_n(i\xi,{\bf q})\right]^2-
4\omega_{\bf q}^2\left|P_a(i\xi,{\bf q})\right|^2} \ , \nonumber\\
D_{\pi a}(\omega=i \xi,{\bf q})&=&
\frac{2\omega_{\bf q}P_a^*(i\xi,{\bf q})}
{\left[\xi^2+\omega_{\bf q}^2+2\omega_{\bf q}P_n(i\xi,{\bf q})\right]^2-
4\omega_{\bf q}^2\left|P_a(i\xi,{\bf q})\right|^2} \ .
\end{eqnarray}
As already discussed in section \ref{Neel}, a  necessary condition
for the stability of the system is the absence of poles in the Green's functions
at negative
$\omega^2$ (positive $\xi^2$). The most dangerous is the regime of very low frequencies
and momenta ($\omega \ll \epsilon_F$, $q \ll p_F \sim \sqrt{\delta}$). Therefore the
criterion of stability is:
\begin{equation}
\label{stab}
\left[\omega_{\bf q}^2+2\omega_{\bf q}P_n(0,{\bf q})\right]^2
> 4\omega_{\bf q}^2\left|P_a(0,{\bf q})\right|^2
\end{equation}
Using Eq. (\ref{pn}) the stability criterion  can be rewritten as:
\begin{equation}
\label{stab1}
1> \frac{2Z^2t^2}{\pi\sqrt{\beta_1\beta_2}}
\left|1+\frac{\pi\sqrt{\beta_1\beta_2}\rho_s}{Z^2t^2}\right|
+\frac{2Z^2t^2}{\pi\sqrt{\beta_1\beta_2}}
\left|1-\frac{\pi\sqrt{\beta_1\beta_2}\rho_s}{Z^2t^2}\right| \ .
\end{equation}
According to Eq. (\ref{dE2}) the expression 
$1-\pi\sqrt{\beta_1\beta_2}\rho_s/Z^2t^2$ is   negative in the $(1,0)$ phase
since $E_{(1,0)}$ is positive.
Therefore the criterion (\ref{stab1}) reads: 
\begin{equation}
\label{stab2}
1 -4\rho_s =0.28 > 0 \ ,
\end{equation}
meaning that the spiral phase is always stable.
It is important to note that the phase is stable only due to spin
quantum fluctuations, and the stability is an  order from disorder effect.
Without account of the fluctuations (semiclassical approximation),
$\rho_s=1/4$, and hence the effective transverse stiffness vanishes,
$1-4\rho_s=0$. In this case the system becomes marginal, or in essence unstable,
 as pointed out in  Refs. \cite{ShSi2,CM}.
Thus the spin quantum fluctuations make the (1,0) spiral phase stable.
We stress that the  spin quantum fluctuation effects in an undoped antiferromagnet
are known quite accurately from numerous approaches
(spin-wave, series expansions, Monte-Carlo, etc.).
Within the chiral perturbation theory, used in the
present work,  one can express the leading terms  in the hole density  
via the  parameters of the undoped system. In this sense we perform an
``exact'' account of the quantum fluctuations.
However, we do not calculate terms subleading in the hole density 
because such a calculation cannot be performed without uncontrolled
mean field approximations. It is possible that at sufficiently
large doping the spiral phase becomes unstable due to such subleading
terms. One  possible  type of  instability is towards formation of a
 noncoplanar state, 
as suggested in \cite{CM}. 
We cannot determine the exact value of doping $\delta$ which is 
  small enough to justify our calculations, but we claim
that at sufficiently small $\delta$ our approach is parametrically justified,
 and consequently 
 Fig. \ref{Fig5} represents the  phase diagram of the
$t-t'-t''-J$ model. We emphasize that
 in this regime the   (1,0) spiral phase  is stable  
 for values of parameters corresponding to
real cuprates.

\section{On-site magnetization in the spiral state}\label{staggered}
\begin{figure}[ht]
\centering
\includegraphics[height=160pt,keepaspectratio=true]{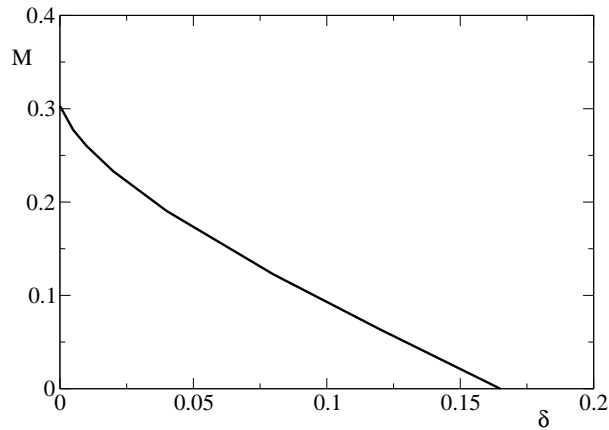}
\caption{\it On-site magnetization in the (1,0) spiral versus doping for
$t=3.1$, $t'=-0.8$, and $t''=0.7$.} 
\label{Fig7a} 
\end{figure}

In the undoped t-J model the spins order in a staggered collinear pattern
 and the value of the staggered
magnetization is:
\begin{equation}
\label{stag1}
M=|\langle S^z \rangle|\approx 0.303 \ .
\end{equation}
In a spiral state the on-site magnetization 
follows a spiral pattern
and besides that the 
value of the magnetization is reduced compared
to (\ref{stag1}). 
We calculate now the reduction of the magnetization.
Using the spin-wave operators (\ref{swt}) and neglecting corrections
proportional to $Q^2 \propto \delta^2$
one can rewrite the on-site magnetization in the following form:
\begin{eqnarray}
\label{stag2}
M=\frac{1}{2}\langle S_a^z-S_b^z \rangle &=&0.303-
2\sum_{\bf q}\frac{1}{\omega_{\bf q}}\langle
[\alpha_{\bf q}^{\dag}\alpha_{\bf q}+
\beta_{\bf q}^{\dag}\beta_{\bf q}-\gamma_{\bf q}(
\alpha_{\bf q}\beta_{\bf -q}+
\alpha_{\bf q}^{\dag}\beta_{\bf -q}^{\dag})]
\rangle \nonumber\\
&=&
1-2\sum_{\bf q}\frac{1}{\omega_{\bf q}}\langle
[\alpha_{\bf q}^{\dag}\alpha_{\bf q}+
\beta_{\bf -q}\beta_{\bf -q}^{\dag}-\gamma_{\bf q}(
\alpha_{\bf q}\beta_{\bf -q}+
\alpha_{\bf q}^{\dag}\beta_{\bf -q}^{\dag})]
\rangle \ .
\end{eqnarray}
For  small doping the deviation of $M$ from the value (\ref{stag1}) 
is due to quantum fluctuations at small momenta, $q \ll 1$. Hence one can 
replace  $\gamma_{\bf q} \to 1$ and rewrite (\ref{stag2}) in terms of  
$\pi$-field  averages  and then  the normal Green's function 
Eqs. (\ref{pi}), (\ref{D}):
\begin{eqnarray}
\label{stag3}
M&\to&
1-2\sum_{\bf q}\frac{1}{\omega_{\bf q}}\langle\pi_{\bf q}^{\dag}\pi_{\bf q}
\rangle = 1-2i\sum_{\bf q}\frac{1}{\omega_{\bf q}}D_{\pi n}(t=-0,{\bf q})\ .
\end{eqnarray}
Finally, using the explicit expression (\ref{Dna}) for the Green's function
we find the reduction of the on-site  magnetization in a spiral:
\begin{eqnarray}
\label{dM}
M&=&0.303+\Delta M \ ,\nonumber\\
\Delta M&=&-4\sum_{\bf q}\int \frac{d\xi}{2\pi}\left\{
\frac{\xi^2+\omega_{\bf q}^2+2\omega_{\bf q}P_n(i\xi,{\bf q})}
{\left[\xi^2+\omega_{\bf q}^2+2\omega_{\bf q}P_n(i\xi,{\bf q})\right]^2-
4\omega_{\bf q}^2\left|P_a(i\xi,{\bf q})\right|^2} 
-\frac{1}{\xi^2+\omega_{\bf q}^2}\right\} \ .
\end{eqnarray}
A straightforward numerical integration of this equation with parameters 
corresponding to Eq. (\ref{ts}) gives the on-site magnetization plotted in
Fig. \ref{Fig7a}.
 
 For  hole concentration 
 $\delta=0.12$, corresponding to a commensurate spiral (Fig. \ref{45spiral}),
 the value of the magnetization is
$M\approx0.06$. It is  quite close to the reported experimental values in
La$_{1.88}$Sr$_{0.12}$CuO$_{4}$ \cite{moment} as well as La$_{1.48}$Nd$_{0.4}$Sr$_{0.12}$CuO$_{4}$
\cite{Tranq1}. The overall decrease of $M$ as a function of doping is also consistent
 with experiment \cite{Wakimoto}. 
 We  observe that
 the spiral state disappears at $\delta_{c} \approx 0.16$, beyond which
 point ($\delta>\delta_{c}$) 
spin rotational invariance is restored. The existence of a  magnetic quantum critical
 point inside the superconducting region is  an issue of considerable current interest
\cite{Sach} (the spiral states support superconductivity as discussed
 in Section VI). It has also  been suggested that the phase emerging  upon 
 destruction of  non-collinear  magnetic order could be of RVB type
 and exhibit electron fractionalization \cite{Sach1,Sach}, although  in the present work 
we can  not address the structure of  the magnetically disordered region $\delta>\delta_{c}$.

\vspace{0.5cm}
\noindent

Using the expression for the polarization operator (\ref{pn})
one can show that $\Delta M$ can be expanded in powers of $\delta$ 
in the following way:
\begin{equation}
\label{dM1}
\Delta M = -a\delta\ln(1/\delta) +b\delta + c\delta^2 +... \ .
\end{equation}
This equation shows that the expansion is nonanalytic in $\delta$.
We have calculated the coefficients  $a$ and $b$ (as well as $c$)  numerically for 
several values of 
$t'$ and $t''$, and tabulated them   in Table II.

Only the logarithmic term in the expansion (\ref{dM1}) is
parametrically justified within the chiral perturbation theory which
we use in the present work.
This is due to the fact  that the integral in Eq. (\ref{dM}) behaves as
$\delta \int dq/q$ and  consequently the ``b''-term in (\ref{dM1}) is
related to the upper limit of the integration, i.e. to the
short distance dynamics ($q\sim \pi$).
\begin{center}
\begin{tabular}{ccccccc}\hline
spiral state & \ \ \ \ $t'$ \ \ \ \  & \ \ \ \ $t''$ \ \ \ \ &  \ \ \ \ &  &
\ \ \ \ a \ \ \ \ & \ \ \ \ b \ \ \ \ \\
\hline
(1,0)   & -0.8 & 0.7  & & & 1.3 &  -1.9 \\
(1,0)  & -0.45 & 0.45 & & & 1.2 & -1.3 \\
(1,0)  & -0.3 & 0.25 & & & 1.0 & -0.7 \\
\hline
\end{tabular}\\
\vspace{10pt}
Table II. {\it Coefficients $a$ and $b$ in the $\delta$-expansion
(\ref{dM1}) for the reduction of the on-site magnetization
for different values of the hopping parameters $t'$ and $t''$.}
\vspace{10pt}
\end{center}
Eq. (\ref{dM}) overestimates the contribution of large momenta
because it does not take into account the reduction of the quasiparticle
residue away from ${\bf k_0}=(\pm \pi/2,\pm\pi/2)$. On the other hand
it underestimates this  contribution since it does not take into account
the incoherent part of the hole Green's function.
To estimate the uncertainty related to the short distance dynamics
we have calculated the reduction of the
nearest-neighbor sites spin-spin 
correlator $\langle {\bf S}_i\cdot {\bf S}_j\rangle$
in the same way, i.e. assuming no variation of the quasiparticle residue and 
without the
incoherent contribution. The result  of this calculation agrees within 30\%
with the available numerical data \cite{HP}. This indicates that the values of the
coefficient ``b'' in Table II are somewhat reliable.
We stress once again that the leading ``a''-term in (\ref{dM1}) is
due to the long distance dynamics only and therefore it is
{\it parametrically} justified and  reliable.

\section{Superconducting pairing in the spiral due to  spin-wave
exchange}
\begin{figure}[ht]
\centering
\includegraphics[height=70pt,keepaspectratio=true]{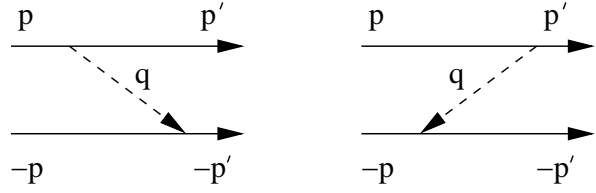}
\caption{\it Spin-wave exchange between holes leading to superconducting
pairing} 
\label{Fig7} 
\end{figure}

Now we investigate  the possibility of
superconducting pairing  due to the spin-wave exchange shown in 
Fig. \ref{Fig7}.
We consider pairing between the  states  $\psi_{{\bf p}}$, Eq. (\ref{psis}),
inside one full pocket (see the right hand side 
in Fig. \ref{Fig3} for the (1,0) state).  
The corresponding many-body wave function has the form:
\begin{equation}
\label{BCS}
|\Psi\rangle = \prod_{\bf p}\left(U_{\bf p} +V_{\bf p}
\psi_{{\bf p}}^{\dag}\psi_{{\bf -p}}^{\dag}\right)|0\rangle \ ,
\end{equation}
and represents pairing of spinless fermions.
Within the  full pocket description the typical momentum transfer in the diagrams 
Fig. \ref{Fig7} is ${\bf q}={\bf p - p'}$,
$q =|{\bf q}| \sim p_F \propto \sqrt{\delta} \ll 1 $. In the    
description with half pockets (left hand side in Fig. \ref{Fig3}) 
the momentum transfer is close to the antiferromagnetic
vector $q \approx {\bf G}=(\pm \pi,\pm\pi)$.
\noindent
We use the full pocket description which is more convenient, and
in fact 
Eq. (\ref{BCS}) already assumes such a description with
$V_{\bf -p}=-V_{\bf p}$. Note that this condition does not mean that
we consider  negative parity pairing. 
 Since parity is defined in the
full magnetic Brillouin zone, in order  to consider parity one needs to transfer 
back from the full-pocket description to the half-pocket description.
Such a transition is related to  translation by the
antiferromagnetic vector ${\bf G}$.
The wave function changes sign under such translation,
$V_{\bf k}=-V_{\bf k+G}$. Therefore $V_{\bf -p}=-V_{\bf p}$ implies
that $V_{\bf -k}=+V_{\bf -k}$ and the pairing has positive parity.
 For a discussion of this peculiar
symmetry property we refer the reader to Refs. \cite{KS,FKS}.

Thus  in the pairing channel the typical momentum transfer is
$q \sim p_F \propto \sqrt{\delta}$.
At such momentum transfer the spin-wave Green's function (\ref{Dna}) is 
substantially different from the bare one (\ref{D0}).
Its spectral weight $Im D_{\pi}(\omega)$ is shifted from 
$\omega=\omega_{\bf q}= \sqrt{2}q \propto \sqrt{\delta}$ down to  
$\omega\sim \epsilon_F \propto \delta$.
Unfortunately in this regime we cannot find an analytical solution for
the pairing. Even at small $\delta$ one needs to solve the Eliashberg equations
numerically and  this is outside of the scope of the present work.
However we can put analytically  a lower  limit on the pairing and study
its symmetry. To do so we replace
the renormalized Green's function (\ref{Dna}) in the diagrams Fig. \ref{Fig7}
by the bare Green's function (\ref{D0}). We call this approximation
the ``bare approximation'' - it is similar to the one used in   Ref. \cite{FKS}.
The typical energy transfer in the diagrams of Fig. \ref{Fig7} is
$\omega \sim \epsilon_F \propto \delta \ll \omega_{\bf q} \propto 
\sqrt{\delta}$. Therefore when calculating the diagrams Fig. \ref{Fig7},
one can set $\omega=0$ in the spin-wave Green's function (\ref{D0}).
This means that in the ``bare  approximation'' it is sufficient to
consider only the static interaction.
By calculating the  diagrams Fig. \ref{Fig7} with account of Eqs (\ref{D0}),
(\ref{qppi}), and (\ref{psis}) we find the effective pairing potential
due to the spin-wave exchange:
\begin{equation}
\label{paird}
V^{(dir)}_{\bf p,p'}=-\frac{M_{\bf q}^2}{\omega_{\bf q}}=-8Z^2t^2\frac{q_1^2}{q^2}
=-8Z^2t^2\frac{(p_1-p'_1)^2}{({\bf p-p'})^2} \ .
\end{equation}
Recall that $p_1$ is the  component orthogonal to the face of the
magnetic Brillouin zone.
Note that in agreement with the general Adler relation \cite{Adler} 
$M_{\bf q} \to 0$,  $q \to 0$ and hence the pairing potential
remains finite  at $q=0$, as  was pointed out by Schrieffer \cite{Sch}.
 In addition to the  diagrams in Fig. \ref{Fig7} there are
also exchange diagrams which differ by  permutation of the legs.
The total effective pairing potential due to the spin-wave exchange is
 then:
\begin{equation}
\label{pair}
V_{\bf p,p'}=V^{(dir)}-V^{(exch)}
=8Z^2t^2\left[-\frac{(p_1-p'_1)^2}{({\bf p-p'})^2}
+\frac{(p_1+p'_1)^2}{({\bf p+p'})^2}\right] \ .
\end{equation}

The BCS equation for the  superconducting gap 
$\Delta_{\bf p}=-\Delta_{\bf -p}$ reads:
\begin{equation}
\label{BCSe}
\Delta_{\bf p}=-\frac{1}{4}\sum_{\bf p'}V_{\bf p,p'}\frac{\Delta_{\bf p'}}
{E_{\bf p'}}=
-\frac{1}{2}\sum_{\bf p'}V_{\bf p,p'}^{(dir)}\frac{\Delta_{\bf p'}}
{E_{\bf p'}} \ ,
\end{equation}
where $E_{\bf p}=\sqrt{(\epsilon_{\bf p}-\mu)^2+\Delta_{\bf p}^2}$.
Note that using the interaction (\ref{pair}) one has to put the coefficient
1/4 instead of 1/2 in the BCS equation to avoid double counting in the
wave function (\ref{BCS}). Alternatively one can use the standard form
with 1/2 and with the direct interaction (\ref{paird}) only. 
In the weak-coupling  limit 
the gap is small compared to the Fermi energy, therefore  Eq. (\ref{BCSe})
can be solved in the vicinity of the Fermi surface 
($\beta_1p_1^2/2+\beta_2p_2^2/2\approx \epsilon_F$)
with logarithmic accuracy. The solution is  discussed in Ref. \cite{FKS}
and we directly present the result: 
\begin{eqnarray}
\label{sol}
\Delta_{\bf p}&=&\Delta_{SC} \sin m\phi \ , \nonumber\\
\Delta_{SC}&=&C\epsilon_Fe^{-1/g_m} \ , \nonumber\\
g_m&=&\frac{4Z^2t^2}{\pi\beta_2(\beta_1/\beta_2-1)}
\left(\frac{\sqrt{\beta_1/\beta_2}-1}{\sqrt{\beta_1/\beta_2}+1}\right)^m \ .
\end{eqnarray}
\begin{figure}[ht]
\centering
\includegraphics[height=150pt,keepaspectratio=true]{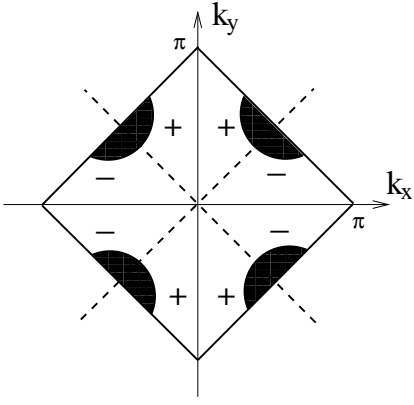}
\caption{\it
d-wave like symmetry of the superconducting gap in the (1,0) spiral state.}
\label{gapsymmetry}
\end{figure}
Here $\Delta_{SC}$ is the maximum value of the superconducting gap. The  
angle $\phi$ is defined as:
\begin{equation}
\label{fi}
\sin\phi=\frac{\sqrt{\beta_2}p_2}{\sqrt{\beta_1p_1^2+\beta_2p_2^2}} \ ,
\end{equation}
$m=1,2,3,..$ is an integer number, and $C\sim 1$ is a constant.
 Eq.  (\ref{sol}) represents a family
 of solutions, however the pairing is maximum in the channel with $m=1$.
In this case, near the Fermi surface, the gap has the simple form
(with $p_1$ and  $p_2$ defined in Fig. \ref{Fig1}):
\begin{equation}
\label{m1}
\Delta_{\bf p} =  \Delta_{SC} \frac{p_2}{\sqrt{(\beta_1/\beta_2)p_1^2+p_2^2}}.
\end{equation}
We observe that   there is a line of nodes 
  along the $(1,1)$ direction. 
  
 The solution (\ref{sol}) only describes the pairing 
within one full pocket while for a (1,0) spiral (which is of primary interest in the
present work) there are two full pockets
(Fig. \ref{Fig3}). In this case one needs to solve the BCS equation numerically
(i.e. take into account the pocket-pocket scattering) in order to determine the symmetry
of the  solution.
This has been done for a similar problem  in  Ref. \cite{BCDS}
and the  expected result is shown in Fig. \ref{gapsymmetry}, where we also
have gone back to the full Brillouin zone (half pocket) description. 
Since the symmetry of the lattice is  spontaneously broken by the spiral,
the ``d-wave'' classification has lost meaning.  
 In principle this should be reflected as  an asymmetry of the pockets themselves;
this asymmetry is small for $\delta \ll 1$ and is 
 not shown in Fig. (\ref{gapsymmetry}). 

\begin{figure}[ht]
\centering
\includegraphics[height=70pt,keepaspectratio=true]{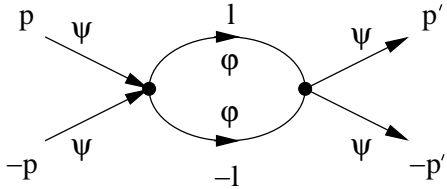}
\caption{\it Second order correction to the pairing interaction between
the $\psi$-fermions due to virtual excitations to the empty $\varphi$-band.
The dot represents the spin-wave exchange shown in Fig. \ref{Fig7}} 
\label{Fig8} 
\end{figure}

The pairing (\ref{sol}) obtained in the ``bare  approximation''
is rather weak.
 As we have already mentioned
 the use of the exact spin-wave Green's function (\ref{Dna})
is expected to  enhance the pairing substantially  due to the spectral
weight shift towards low frequencies; we plan to discuss the
full numerical solution of the Eliashberg equations in a future work. 
However there exists an additional contribution
to pairing (not accounted for in the solution (\ref{sol})) which
can be estimated already in the weak-coupling limit. 
This  involves pairing via the upper $\varphi$-band
(ghost band) and can be included by replacing 
 the many-body wave function (\ref{BCS})
 by:
\begin{equation}
\label{BCS1}
|\Psi\rangle = \prod_{\bf p}\left(U_{\bf p} +V_{\bf p}
\psi_{{\bf p}}^{\dag}\psi_{{\bf -p}}^{\dag}\right)
\prod_{\bf p'}\left(\overline{U}_{\bf p'} +\overline{V}_{\bf p'}
\varphi_{{\bf p'}}^{\dag}\varphi_{{\bf -p'}}^{\dag}\right)|0\rangle \ .
\end{equation}
To account for this effect we calculate the second order correction
to the spin-wave mediated interaction (\ref{paird}) between the 
$\psi$-fermions. The correction is given by
the diagram in Fig. \ref{Fig8}.
\noindent 
The dot in this diagram is given by Fig. \ref{Fig7} where  the ``flavor''
of the fermion is changed from $\psi$ to $\varphi$ in each spin-wave vertex.
Calculating the diagrams in Fig. \ref{Fig7} with account of Eqs. (\ref{D0}),
(\ref{qppi}), and (\ref{psis}) we find that the dot is described by the same 
Eq. (\ref{paird}) but with a sign minus: $''dot''=M_{\bf q}^2/\omega_{\bf q}
=8Z^2t^2(p_1-l_1)^2/({\bf p-l})^2$. 
Hence the second order correction Fig. \ref{Fig8} reads:
\begin{eqnarray}
\label{2nd}
\delta V^{(dir)}_{\bf p,p'}=-(8Z^2t^2)^2\sum_{\bf l}
\frac{(p_1-l_1)^2}{({\bf p-l})^2} \
\frac{1}{2\epsilon_{\bf l}+2\Delta-2\epsilon_F} \
\frac{(p'_1-l_1)^2}{({\bf p'-l})^2} \ ,
\end{eqnarray}
where $\epsilon_{\bf l}$ is given by Eq. (\ref{e}), and $\epsilon_F$ and
$\Delta$ are given by Eqs. (\ref{ef10}), (\ref{dd})
for the  $(1,0)$ spiral state.
The evaluation of the sum in Eq. (\ref{2nd}) is particularly simple for the
isotropic case, $\beta_1=\beta_2$. We will  consider this case
for the purpose of an estimate of the effect.
Using the expansion with Chebyshev polynomials $T_n(x)$:
\begin{equation}
\frac{1}{1-2tx+t^2}=\frac{1}{1-t^2}\left(1+2\sum_{n=1}^{\infty}t^nT_n(x)\right)
 \ ,
\end{equation}
and keeping only the first angular harmonic which is important for
pairing in the $m=1$ channel, we obtain:
\begin{equation}
\frac{(p_1-l_1)^2}{({\bf p-l})^2} \to 
-\frac{1}{2}\ \frac{p_<}{p_>}\cos(\phi_p+\phi_l) \ ,
\end{equation}
where $p_<=min(p,l)$ and  $p_>=max(p,l)$.
Using this representation and having in mind that in Eq. (\ref{2nd}) 
$|p|=|p'|=p_F$ we perform the integration in (\ref{2nd}) and find
the  correction
to the pairing interaction:
\begin{eqnarray}
\label{2nd1}
\delta V^{(dir)}_{\bf p,p'}=
-\frac{(8Z^2t^2)^2}{32\pi \beta}\left(1-\frac{\Delta-\epsilon_F}{\epsilon_F}
\ln\frac{\Delta}{\Delta-\epsilon_F}
+\frac{\epsilon_F}{\Delta-\epsilon_F}
\ln\frac{\Delta}{\epsilon_F}\right)
\times(\cos \phi_p\cos\phi_{p'}+\sin \phi_p\sin\phi_{p'}) \  .
\end{eqnarray}
We remind the reader that the above formula is
 valid in the  isotropic case, $\beta=\beta_1=\beta_2$.
The correction (\ref{2nd1}) gives rise to an approximately
 10\% enhancement of the
superconducting coupling constant $g_1$,  Eq. (\ref{sol}):
\begin{equation}
\label{dg}
\frac{\delta g_1}{g_1}=\frac{Z^2t^2}{2\pi \beta}
\left(1-\frac{\Delta-\epsilon_F}{\epsilon_F}
\ln\frac{\Delta}{\Delta-\epsilon_F}
+\frac{\epsilon_F}{\Delta-\epsilon_F}
\ln\frac{\Delta}{\epsilon_F}\right) \ .
\end{equation}
This enhancement is an interesting qualitative effect related to the
``ghost'' band, but   numerically  its influence on pairing is relatively weak.
Taking into account both Eq. (\ref{sol})  and Eq. (\ref{dg}) we obtain
 the following estimate for the pairing in the dominant, $m=1$ channel
 (we use $\beta_{1} = 2.9, \beta_{2}=3.8$ corresponding
to the hoppings from Eq. (\ref{ts}), and  also reinstate the  Heisenberg exchange  $J$):
\begin{equation}
\label{estimate}
 \Delta_{SC} \sim 10^{-2} \delta J.
\end{equation}
 For ten percent doping this weak-coupling formula
produces a $T_c \propto \Delta_{SC}$ of the order of several Kelvin which is 
about 10 times  smaller than  transition temperatures in real compounds.
The most important pairing  enhancement  is expected from the renormalization
of the spin-wave Green's function, to be discussed in a separate work.  

\section{Conclusions and discussion}
We have studied  the phase diagram of the $t-t'-t''-J$ model close to
half-filling. 
To determine the  single  hole properties (one hole injected in an 
antiferromagnetic background) we use  the self consistent Born approximation.
This approximation is not parametrically justified for the $t-J$ model,
but it is known that it works remarkably well for the single hole properties. 
The  crucial observation is that these properties are not related to the long range 
dynamics: all integrals in momentum space are convergent at large momenta
(quantum fluctuations at distances 1-2 lattice spacings).
Finite doping brings nontrivial long-range dynamics into the problem.
To determine the  dynamics we use  the chiral perturbation theory
which involves an  expansion in powers of doping near  half filling,
$\delta \ll 1$. The  efficiency of the chiral perturbation theory is
in essence a consequence of the  dimensionality of the problem (2+1 D) and 
  Goldstone's theorem. 
We certainly cannot determine  the exact value of doping so that it
is small
 enough 
   to justify the approach, but we claim
that at sufficiently small $\delta$ the approach is parametrically justified.
We show that the Neel state is unstable with respect to  decay to spiral states
as soon as doping is introduced.
 The analysis of the stability of the spiral phases is 
performed within the  RPA approximation, and  all non-RPA corrections are
suppressed by powers of $\delta$. 

We find that at $t'=t''=0$ the spiral state is unstable  toward a local 
enhancement of the spiral pitch, and consequently at that point 
 the nature of the true ground state remains 
unclear. However we have shown that  for
 values of $t'$ and $t''$ corresponding to real cuprates
the (1,0) spiral state is stable. 
  The phase diagram of the model at 
$t=3.1$ (we set $J=1$) is shown in Fig. \ref{Fig5}.
For hole concentration
 $\delta\approx 0.119$ the (1,0) spiral is commensurate with the lattice 
with a period of 8 lattice spacings, in agreement with experimental data.
Fig. \ref{Fig7a} shows the dependence of the spiral on-site magnetization
on doping.
We  demonstrate that  spin-wave mediated superconductivity
is developed above the spiral state and
derive analytically a lower limit for the superconducting gap.
One cannot classify the gap according to the lattice
representations A$_1$, A$_2$, B$_1$, B$_2$, and E (``s'',''d'',''p'',...) 
since the   symmetry of the lattice is spontaneously 
broken by the spiral. However, the gap always has lines of nodes
 and  a symmetry similar to d-wave. 

Finally we briefly discuss the possible connection of our results to
 experiments in the cuprates. Magnetic order has been observed in the
 superconducting states of both La$_{2-x- \delta}$Nd$_x$Sr$_{\delta}$CuO$_{4}$
\cite{Tranq,Tranq1}
 and La$_{2- \delta}$Sr$_{\delta}$CuO$_{4}$ \cite{Jul}. This order seems to
 be particularly enhanced near the hole concentration $\delta=1/8$ 
where
a  commensurate  structure with a period of 8 lattice
 spacings is observed. In addition, static charge order (with a period 4) has been observed
at least in one of the materials \cite{Tranq,Tranq1}. This has lead to proposals
 that the state near  $\delta=1/8$ is  a collinear spin density wave (SDW)  coexisting
 with a charge density wave (CDW) at  twice the wave vector \cite{Sach}. Such an interpretation
 of experiments rules out a non collinear spiral configuration 
of the type shown in Fig. \ref{45spiral}, since the density of holes is expected to
 be uniform in the spiral case. On the other hand the question whether (static) charge order
 is a generic feature of the cuprates is far from being resolved and experiments
 in many cuprates are interpreted as showing fluctuating (i.e. dynamic) order in the charge
 sector \cite{Kiv}. For example the existence of (static) charge order in 
 La$_{2- \delta}$Sr$_{\delta}$CuO$_{4}$ 
is far less clear \cite{Jul} although it also exhibits commensurate magnetic peaks around $\delta=1/8$.

In the light of the above we find   our result that the spiral (1,0) state is commensurate 
with the lattice
(period 8, Fig. (\ref{45spiral})) for $\delta=0.119$ very promising. 
  It seems to be consistent with the data in the  magnetic sector 
(although the interpretation of Ref. \cite{Tranq1} rules out a spiral state in that material).
Moreover, it is known that doping introduces disorder, in turn leading to glassy features
 in the magnetism; at least at low doping (in the normal state) those features
  can be explained well within the spiral scenario \cite{HCS}.   
In our opinion it would be also  very interesting to study the density response of the system
 in the spiral state. In the context of the $t-J$ model the density response has been
recently investigated in the Neel state \cite{Horsch}, where fluctuations at energies of
order $J$ were found. In  a spiral state, due to the presence of the  ghost (empty) band,
we expect that even lower energy charge density fluctuations will exist, although a
perfectly static
CDW order seems impossible to stabilize.  
 It is certainly more likely that  some kind of order in the charge sector
appears (if anywhere at all) around the point $t'=t''=0$, where the spiral state
 is inherently unstable. If we indeed interpret the region marked
 as ``unstable'' on our phase diagram  Fig. \ref{Fig5}  as a candidate for such order, 
then  the effect  of increasing $t'$ and/or $t''$ is to  drive the system towards 
  the (stable) spiral order, with a homogeneous charge distribution. This occurs
for $|t'|/t \gtrsim 0.18$ (at $t''=0$). 
From this point of view our results are similar, at least superficially,
  to the DMRG results
 describing the destruction of 
  stiped phases by  second-neighbor hopping \cite{White, Tohyama},
 where a critical value of $t'$ necessary to destabilize the
 stripes  was found, and is quite close to our estimate above.
\\

We acknowledge discussions with J. Haase, G. Khaliullin, P. Horsch, 
C. Bernhard, and S. Sachdev. An important part of this work was  done 
during the  stay of one of us (O.P.S.) at the  Max-Planck-Institute fur
Festkorperforschung, Stuttgart. He is very grateful to all members of the
theoretical department for hospitality and very stimulating atmosphere.
O.P.S. also thanks the
 Institut de Physique Th\'eorique 
(Universit\'e de Lausanne), where part of the research was performed,
for hospitality 
and acknowledges support from the {\it Fondation Herbette}.


\begin{thebibliography}{99}
\bibitem{And0} P. W. Anderson, Science {\bf 235}, 1196 (1987).
\bibitem{Emery} V. J. Emery, Phys. Rev. Lett. {\bf 58}, 2794 (1987).
\bibitem{ZR} F. C. Zhang and T. M. Rice, Phys. Rev. B {\bf 37}, 3759 (1988).
\bibitem{ShSi} B. I. Shraiman and E. D. Siggia, Phys. Rev. Lett.
{\bf 62}, 1564 (1989).
\bibitem{ShSi2} B. I. Shraiman and E. D. Siggia, Phys. Rev. B
{\bf 46}, 8305 (1992).
\bibitem{Dom} T. Dombre, J. Phys. (France) {\bf 51}, 847 (1990).
\bibitem{Ng} C. L. Kane, P. A. Lee, T. K. Ng, B. Chakraborty, and N. Read,
 Phys. Rev. B {\bf 41}, 2653 (1990).
\bibitem{Auer} A. Auerbach and B. E. Larson, Phys. Rev. B {\bf 43},
7800 (1991).
\bibitem{IF} J. Igarashi and P. Fulde, Phys. Rev. B {\bf 45},
10419 (1992).
\bibitem{Mori} H. Mori and M. Hamada,  Phys. Rev. B {\bf 48}, 6242 (1993).
\bibitem{CM} A. V. Chubukov and K. A. Musaelian, Rev. B {\bf 51}, 12605 (1995).
\bibitem{Bruce} B. Normand and P. A. Lee, Phys. Rev. B {\bf 51}, 15519 (1995).
\bibitem{Manuel} L. O. Manuel and H. A. Ceccatto,  Phys. Rev. B {\bf 61}, 3470 (2000).
\bibitem{Jul} M. -H. Julien, Physica B {\bf 329-333}, 693 (2003).
\bibitem{HCS} N. Hasselmann, A. H. Castro Neto, and C. Morais Smith,
cond-mat/0306721.
\bibitem{White} S. R. White and D. J. Scalapino, Phys. Rev. B {\bf 60}, R753 (1999).
\bibitem{Tohyama} T. Tohyama, C. Gazza, C. T. Shih, Y. C. Chen,
T. K. Lee, S. Maekawa, and E. Dagotto, Phys. Rev. B {\bf 59}, R11649 (1999).
\bibitem{Tranq} J. M. Tranquada, B. J. Sternlieb, J. D. Axe, 
Y. Nakamura, and S. Uchida, Nature {\bf 375}, 561 (1995).
\bibitem{Tranq1} J. M. Tranquada, J. D. Axe, N. Ichikawa, 
Y. Nakamura, S. Uchida, and B. Nachumi, Phys. Rev. B {\bf 54}, 7489 (1996).
\bibitem{ShSi1} B. I. Shraiman and E. D. Siggia,
 Phys. Rev. B {\bf 40}, 9162 (1989).
\bibitem{KS} M. Yu. Kuchiev and O. P. Sushkov, Physica C {\bf 218}, 197 
(1993).
\bibitem{FKS} V. V. Flambaum, M. Yu. Kuchiev, and O. P. Sushkov, Physica C 
{\bf 227}, 267 (1994).
\bibitem{BCDS} V. I. Belinicher, A. L. Chernyshev, 
A. V. Dotsenko, and O. P. Sushkov, Phys. Rev. B {\bf 51}, 6076 (1995).
\bibitem{CP} See, e.g. S. Weinberg, Phys. Rev. Lett. {\bf 17}, 616 (1966);
Phys. Rev. {\bf 166}, 1586 (1968). See also:  
F. Hasenfratz and F. Niedermayer, Z. Phys. B {\bf 92}, 91 (1993) and
references  therein.
\bibitem{Adler} S. L. Adler, Phys. Rev. {\bf 137}, B1022 (1965).
\bibitem{Sch} J. R. Schrieffer, J. Low Temp. Phys. {\bf 99}, 397 (1995).
\bibitem{Tok} Y. Tokura, S. Koshihara, T. Arima,
H. Takagi, S. Ishibashi, T. Ido, and S. Uchida,
 Phys. Rev. B {\bf 41}, 11657 (1990).
\bibitem{Grev} M. Greven, R. J. Birgeneau, Y. Endoh, M. A. Kastner, B. Keimer, 
M. Matsuda, G. Shirane, and T. R. Thurston, 
 Phys. Rev. Lett. {\bf 72}, 1096 (1994).
\bibitem{And} O. K. Andersen, A. I. Liechtenstein, O. Jepsen, and F. Paulsen, 
J. Phys. Chem. Solids {\bf 56}, 
1573 (1995).
\bibitem{SSEE} O. P. Sushkov, G. A. Sawatzky, R. Eder, and H. Eskes,
 Phys. Rev. B {\bf 56}, 11769 (1997).
\bibitem{Schmitt} S. Schmitt-Rink, C. M. Varma, and 
A. E. Ruckenstein, Phys. Rev. Lett. {\bf 60}, 2793 (1988).
\bibitem{Kane} C. L. Kane, P. A. Lee, and N. Read, Phys. Rev. B {\bf 39},
6880 (1989).
\bibitem{Mart} G. Martinez and P. Horsch, Phys. Rev. B {\bf 44}, 317 (1991).
\bibitem{Liu} Z. Liu and E. Manousakis, Phys. Rev. B {\bf 45}, 2425 (1992).
\bibitem{Manousakis} E. Manousakis, Rev. Mod. Phys. {\bf 63}, 1 (1991).
\bibitem{SF}  O. P. Sushkov and V. V. Flambaum, Physica C {\bf 206}, 269 
(1993).
\bibitem{MS} D. Murray and O. P. Sushkov, Physica C {\bf 258}, 389 (1996).
\bibitem{SZ} R. R. P. Singh, Phys. Rev. B {\bf 39}, 9760 (1989);
Zheng Weihong, J. Oitmaa, and C. J. Hamer, Phys. Rev B
{\bf 43}, 8321 (1991).
\bibitem{Eder} R. Eder, Phys. Rev. B {\bf 43}, 10706 (1991).
\bibitem{moment}  H. Kimura, H. Matsushita, K. Hirota,  Y. Endoh,
K. Yamada, G. Shirane, Y. S. Lee, M. A. Kastner, and R. J. Birgeneau,
Phys. Rev. B {\bf 61}, 14366 (2000).
\bibitem{Wakimoto} S. Wakimoto, R. J. Birgeneau, Y. S. Lee, and G. Shirane,
Phys. Rev. B {\bf 63}, 172501 (2001). 
\bibitem{Sach} S. Sachdev, Rev. Mod. Phys. {\bf 75}, 913 (2003) and references
therein.
\bibitem{Sach1} S. Sachdev, Annals of Physics {\bf 303}, 226 (2003).
\bibitem{HP} Y. Hasegawa and D. Poilblanc, Phys. Rev. B
{\bf 40}, 9035 (1989).
\bibitem{Kiv} S. A. Kivelson, I. P. Bindloss, E. Fradkin, V. Oganesyan, J. M.
Tranquada, A. Kapitulnik, and C. Howald, 
Rev. Mod. Phys. {\bf 75}, 1201 (2003).
\bibitem{Horsch} P. Horsch, G. Khaliullin, and V. Oudovenko,
 Physica C {\bf 341-348}, 117 (2000). 

\end{thebibliography}
\end{document}